\documentclass[12pt,preprint]{aastex}

\newcommand{\etal}{et~al.\ }

\newcommand{\cmsq}{\hbox{cm$^{-2}$}}

\newcommand{\flux}{\hbox{erg~cm$^{-2}$~s$^{-1}$}}

\newcommand{\nh}{\hbox{${N}_{\rm H}$}}

\newcommand{\chandra}{{\emph{Chandra}}}

\newcommand{\xmm}{\emph{XMM-Newton}}
\newcommand{\asca}{{\emph{ASCA}}}

\newcommand{\pks} {PKS~1830$-$211}

\begin{document}

\def\sarc{$^{\prime\prime}\!\!.$}
\def\arcsec{$^{\prime\prime}$}
\def\arcmin{$^{\prime}$}
\def\degr{$^{\circ}$}
\def\seco{$^{\rm s}\!\!.$}
\def\ls{\lower 2pt \hbox{$\;\scriptscriptstyle \buildrel<\over\sim\;$}} 
\def\gs{\lower 2pt \hbox{$\;\scriptscriptstyle \buildrel>\over\sim\;$}} 
 
\title{Variable X-ray Absorption toward Gravitationally-Lensed Blazar \pks}

\author{Xinyu Dai\altaffilmark{1}, Smita Mathur\altaffilmark{1}, George Chartas\altaffilmark{2}, Sunita Nair\altaffilmark{3}, and Gordon P. Garmire\altaffilmark{2}}

\altaffiltext{1}{Department of Astronomy,
The Ohio State University, Columbus, OH 43210, USA
xinyu@astronomy.ohio-state.edu, smita@astronomy.ohio-state.edu}
\altaffiltext{2}{Department of Astronomy and Astrophysics,
The Pennsylvania State University, University Park, PA 16802, USA
chartas@astro.psu.edu, garmire@astro.psu.edu.}
\altaffiltext{3}{Astronomy \& Astrophysics Group, 
Raman Research Institute, Bangalore 560080, India,
sunita@rri.res.in.}

\begin{abstract}
We present X-ray spectral analysis of five \chandra\ and \xmm\
observations of the gravitationally-lensed blazar \pks\ from 2000 to 2004.
We show that the X-ray absorption toward \pks\ is variable, and the
variable absorption is most likely to be intrinsic with amplitudes of
$\sim 2\times10^{22}$--$30 \times10^{22} \cmsq$ depending on whether or not the absorber
is partially covering the X-ray source.  
Our results confirm the variable absorption observed previously, 
although interpreted differently, in a sequence of \asca\ observations.
This large variation in the 
absorption column density can be interpreted as outflows from the
central engine in the polar direction, consistent with recent numerical
models of inflow/outflows in AGNs. 
In addition, it could possibly be caused by the interaction between the 
blazar jet and its environment, or the variation from the geometric 
configuration of the jet.
While the spectra can also be fitted
with a variable absorption at the lens redshift, 
we show that this model is unlikely. We also
rule out the simple microlensing interpretation of variability which was
previously suggested.

\end{abstract}


\section{Introduction}

\pks\ \citep{ps88,su90,ja91} consists of two $z_s =2.507$ \citep{li99}
 blazar images separated by 1\arcsec\ and lensed by a $z_l = 0.886$
 \citep{wc96,ge97} spiral galaxy \citep{wi02}.  \pks\ was observed in
 the radio \citep[e.g.,][]{ps88,lovell98}, infrared
 \citep[e.g.,][]{li99,courbin02,wi02}, X-rays
 \citep{mn97,oshima01,de05,dai06}, and Gamma-rays \citep{mattox97}, and
 the spectral energy distribution of \pks\ \citep{de05} shows two
 emission bumps with one in the infrared and the other between the hard
 X-ray and Gamma-ray bands.

The gravitational lens \pks\ is complicated in many aspects, and one of
them is the X-ray absorption.  The X-ray absorption was first detected
by ROSAT \citep{mn97}.  Later ASCA observations \citep{oshima01} show that 
the X-ray absorption is variable; these authors favored a model in which
microlensing was the cause of variability.  Recently, \pks\ was observed
by \chandra\ and \xmm\ for five epochs, which enabled us to study the
nature of the X-ray absorption in detail combining the angular
resolution of \chandra\ and the large collecting area of \xmm.  Since
the two images are resolved by \chandra, we can test the microlensing
model predictions and study the X-ray absorption separately for the two
lensed images.  \citet{dai06} found that the differential absorption at
the lens galaxy between the two images is $\Delta\nh_{B,A} =
1.8^{+0.5}_{-0.6}\times 10^{22}$~\cmsq.  In this paper, we study the
time evolution of the absorption in the system from the five \chandra\
and \xmm\ observations and confirm that the X-ray absorption toward
\pks\ is variable. Moreover, we rule out the microlensing interpretation
of the observed variability.

\section{Observations and Data Reduction}
We  observed \pks\ twice with \chandra\ and three times with \xmm\
from 2000 to 2004.  The details of the observations are listed in
Table~\ref{tab:obslog}.  The \chandra\ data were reduced with the
\verb+CIAO3.2+ software tools provided by the \chandra\ X-ray Center (CXC)
following the standard threads on the CXC website.\footnote{The CXC
website is at http://cxc.harvard.edu/.}  Only events with standard
\asca\ grades of 0, 2, 3, 4, and 6 were used in the analysis.  We
improved the image quality of the data by removing the pixel
randomization applied to the event positions by the standard pipeline.
In addition, we applied a sub-pixel resolution technique \citep{t01,m01}
to the events on the S3 chip of ACIS where the quasar images are
located. This allowed us to resolve the two lensed images of the
blazar. The \xmm\ data were reduced using the standard analysis software
\verb+SAS6.0+.  We used the tasks \verb+epchain+ and \verb+emchain+ from
\verb+SAS+ to reduce the PN and MOS data, and filtered the events with
patterns $\le 4$ and $\le 12$ for the PN and MOS chips, respectively.

\section{Spectral Analysis}
We fitted the spectra of  \pks\ using \verb+XSPEC V11.3.1+
\citep{a96} in the 0.35--8 keV observed energy range for \chandra\
spectra and in the 0.35--10 keV range for \xmm\ spectra.  In all of our
models, we fixed the Galactic absorption at $\nh =
0.22\times10^{22}$~\cmsq\ \citep{d90}.

\subsection{\chandra\ Spectra of Individual Images\label{sec:indi}}
We analyzed the \chandra\ spectra of individual images A and B.  We
modeled the spectra with a power law modified by the Galactic absorption and
the absorption at the lens redshift; the fitting results are listed in
Table~\ref{tab:indi}. We experimented with two models where the
power law photon indices for the two images were allowed to be different
(Model 1) and where they were constrained to be the same (Model 2).  We
assumed that the excess absorption (above Galactic) arises in the lens
galaxy at z=0.886 (we test this assumption later). In both the models,
the column densities at the two \chandra\ epochs for image B are
consistently larger than those of image A.  In addition, the absorption
in the first epoch is consistently larger than that in the second epoch
for both images, except for image B in Model 1 where it is similar
within errors.  The photon indices obtained from Model 1 fits are
similar for the two images and in both epochs, partially due to the
large error bars on the parameter caused by the low signal-to-noise ratio of
the individual spectra.  The photon indices obtained from Model 2 show
$\sim1\sigma$ difference from epoch to epoch, which again is not
significant.  The difference between the absorption in the two images
occurs simply because the lines of sight of the two images intersect
different parts of the lens galaxy \citep{dai06}.  The difference
between the absorption at the two epochs indicates that the X-ray
absorption is variable.  The origin of the absorption variability could
be either at the lens redshift or at the source redshift.  We note
that the differential absorption between images B and A is similar in
the two epochs, especially in Model 2.  This is suggestive of absorption
variability occurring at the source redshift.

\subsection{\chandra\ and \xmm\ Spectra of Combined Images}

We then analyzed the spectra of the combined images AB for the \chandra\
and \xmm\ observations.  We first fitted the spectra of \pks\ of the
five epochs with a power-law model modified by neutral absorption from
the Milky Way and the lens galaxy.  The fitting results are listed in
Table~\ref{tab:spec} (Model 3).  We note that the absorption at the lens
redshift for this model should be treated as an averaged absorption for
the two lines of sight. 
We also fitted the co-added three \xmm\ spectra to obtain
a higher signal-to-noise ratio spectrum, and the results are also listed in 
Table~\ref{tab:spec}. The higher S/N of \xmm\
spectra allows better constraints on the power-law spectral index. We
find that the spectral index varies between the \chandra\ and the \xmm\
observations, though variations within the \chandra\ epochs and 
\xmm\ epochs are smaller.
Comparing the fitting results of Model 3 and Model 2, the spectral index 
obtained by fitting the combined images AB for the \chandra\ observations
are consistent with results from individual spectral fits in
\S~\ref{sec:indi}.  The absorption at the lens is also variable,
especially when comparing the second \chandra\ observation with other
epochs. 
We further test the variability of the spectra by comparing a model with
no spectral variability (with the exception of normalization) and another model
with variable absorption and spectral index.
We found that the model with the spectral variability produced a better fit
with a $\Delta \chi^2$ improvement of 56.1 (a null probability of $10^{-7}$ by the F-test)
when jointly fitting the five X-ray observations. 
When we only consider the three \xmm\ observations, the corresponding 
improvement is 14.5 with a null probability of 0.0097 given by the F-test.
The largest absorption variation is $\Delta \nh = (0.7\pm0.3)\times10^{22}~\cmsq$ between the 
second \chandra\ epoch and the first \xmm\ epoch.
Although the variable spectral index in blazars is common
\citep[e.g.,][]{fos06}, the absorption at the lens galaxy is unlikely to
vary on the time scales of years.  As discussed in \S~\ref{sec:indi},
it is more likely that the variable absorption component is at the
source redshift close to the AGN, where a short time-scale variability is
possible and may be expected.

The next model we tried is a power-law modified by three absorption
components, the Galactic absorption and the absorptions at the lens and
the source.  We fixed the absorption at the lens as $\nh =
1.7\times10^{22}$~\cmsq, the smallest absorption column density detected
in the five epochs from the previous model.  This is the largest
absorption column density at lens we can set to ensure no absorption
variability at the lens galaxy.
The fitting results are listed in Table~\ref{tab:spec} (Model 4).  Again
we detected variability of both the photon index and the absorption at
the source redshift.  
Using the same test that we described in the previous model, we found
that the model allowing spectral variability improved the $\Delta \chi^2$ 
by 59.8 with a null probability of $5\times10^{-8}$ for jointly fitting
the five X-ray observations, and an improvement of 14.7 with a null
probability of 0.01 for jointly fitting the three \xmm\ observations.
The largest absorption variation for this model is $\Delta \nh = (3.5\pm0.7)\times10^{22}~\cmsq$ between the 
second \chandra\ epoch and the first \xmm\ epoch.
The variable absorption in this model can be
naturally associated with outflows from the central engine.  
Although, this model produces comparable fits for the two \chandra\ spectra
compared with Model 3, where
there is no absorption at the source redshift,
the fits for the \xmm\ spectra with a higher S/N are worse than Model 3 with $\Delta \chi^2$
increases of 5.8, 37.2, 13.0 for the three \xmm\ spectra.  
We also tested whether
fitting the absorption component at the source with a warm absorber or a
partially covering absorber would improve the fit.  While the warm
absorber model produces a worse fit, the partial covering model
(Table~\ref{tab:spec}, Model 5) produces comparable fits to Model 3.
For the partial covering model, the model allowing spectral variability
produces improvement of the $\Delta \chi^2$ 
by 79.7 with a null probability of $1\times10^{-11}$ for jointly fitting
the five X-ray observations, and an improvement of 14.1 with a null
probability of 0.01 for jointly fitting the three \xmm\ observations.
The largest absorption variation for this model is $\Delta \nh = (29\pm7)\times10^{22}~\cmsq$, again, 
between the second \chandra\ epoch and the first \xmm\ epoch.

\begin{figure}
\epsscale{1}
\plotone{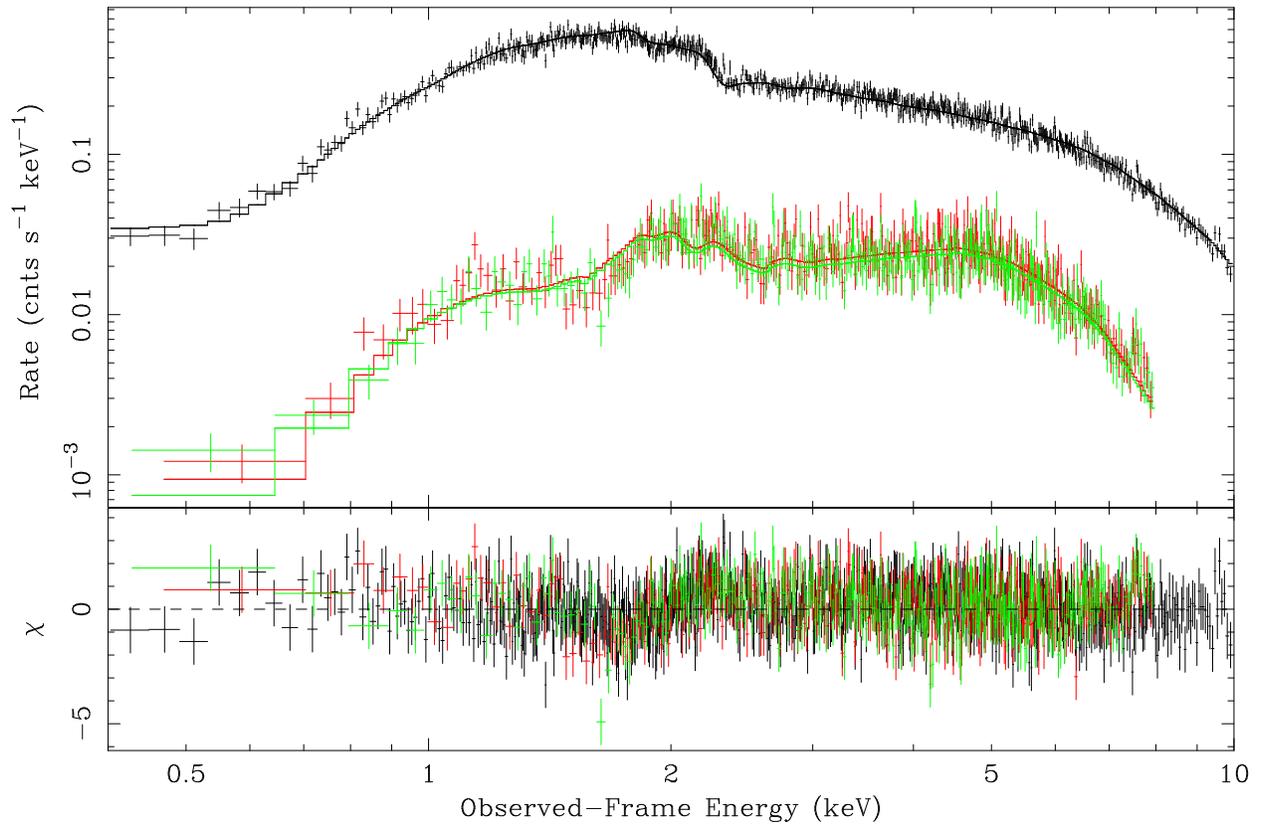}
\caption{\xmm\ and \chandra\ spectra of \pks.  The top spectrum is a
 combined spectrum from three \xmm\ observations, and the bottom two
 spectrum is from the two \chandra\ observations.  The spectra are
 fitted simultaneously with absorbed power-law models as described in
 Model 6 of Table~\ref{tab:spec}. \label{fig:spec}}
\end{figure}

As a final improvement to our model, we assume that the lines
of sight of the two lensed images pass through different locations in
the lens galaxy and thus have different absorption column densities.
This differential absorption is measured as $\Delta\nh_{B,A} =
1.8^{+0.5}_{-0.6}\times 10^{22}$~\cmsq\ \citep{dai06}.  Therefore, we
have a model composed of two power-law components and each of them is
modified by three absorptions from Galactic, lens, and source.  The two
power-law components representing the two lensed images have the same
photon index but different normalizations.  The normalization ratio
between images A and B is constrained as $R = 1.03$ by taking the hard
X-ray flux ratios (3--8 keV) from the \chandra\ observation.  The
Galactic absorption and the absorption at the source are the same for
the two power-law components, and the absorbers at the source are
assumed to be partially covering the continuum.  The absorber column
densities at the lens are different for the two power-law components by
the amount given by \citet{dai06}.  We fit the two \chandra\ spectra and
the co-added \xmm\ spectrum simultaneously with this model.  We
constrained that the absorption at lens did not vary.  The fitting
results are listed in Table~\ref{tab:spec} (Model 6), and the spectra are shown
in Figure~\ref{fig:spec}.  The fitting
results indicate that the absorption at the source has varied by roughly
$(39\pm7) \times10^{22}\cmsq$.

\section{Discussion}

\begin{figure}
\epsscale{1}
\plotone{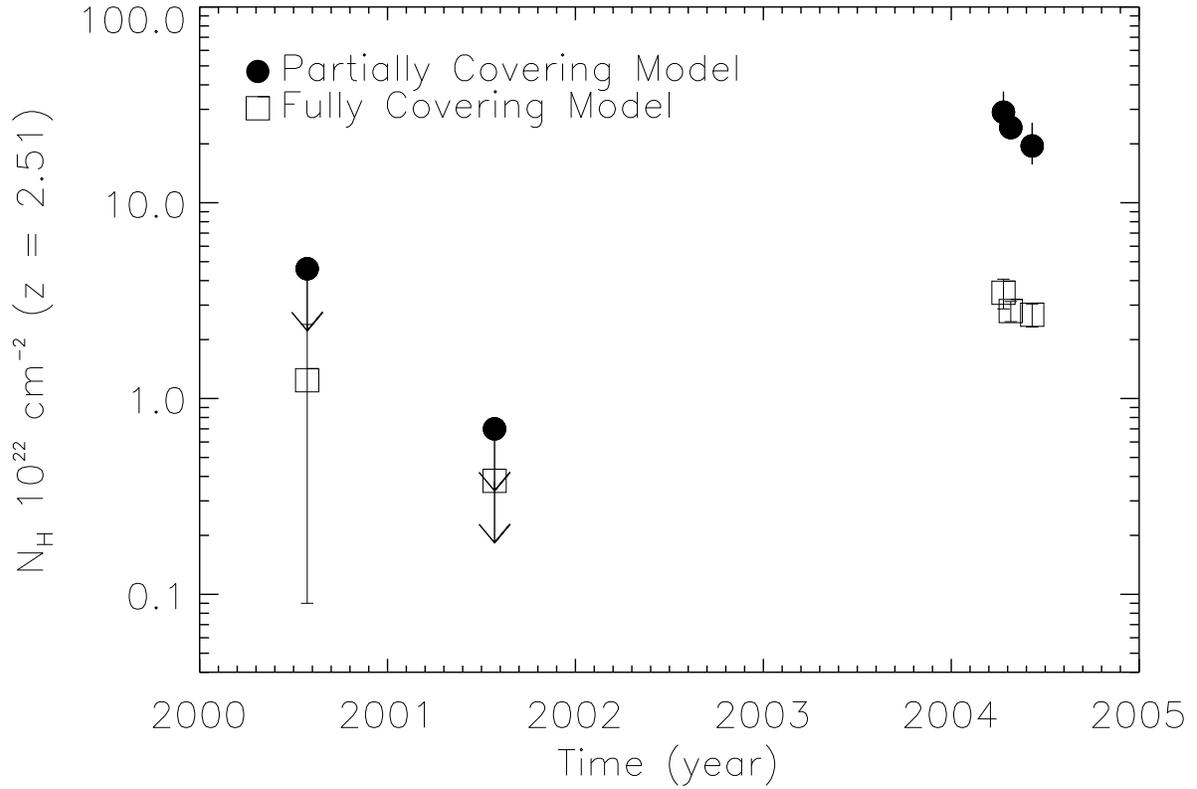}
\caption{The absorption column densities at the redshift of the source ($z=2.51$) of \pks\ measured from \chandra\ and \xmm\ spectra assuming a partial covering model (filled circles) and a fully covering absorption model (squares). \label{fig:nh}}
\end{figure}

\begin{figure}
\epsscale{1}
\plotone{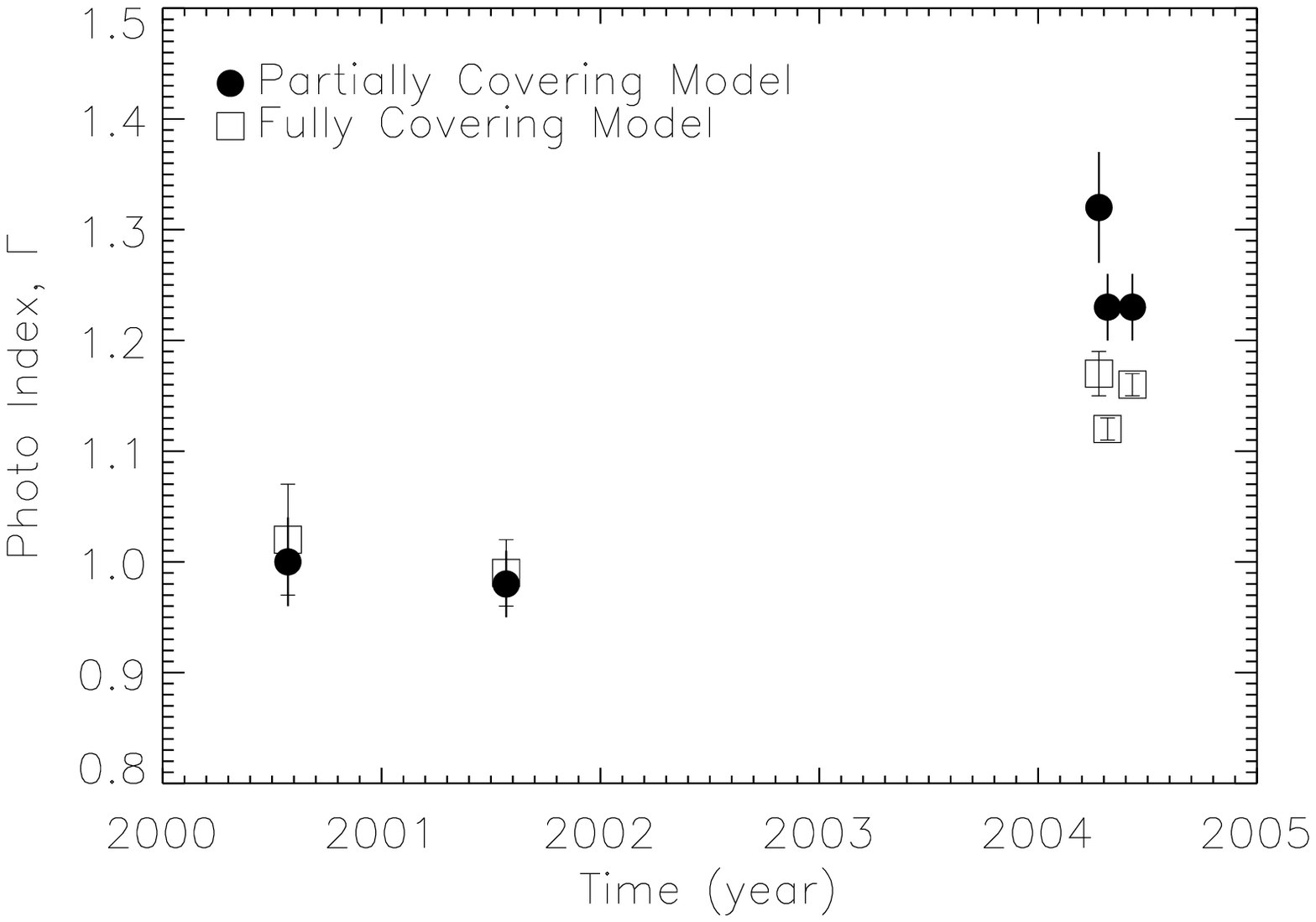}
\caption{The photon indices measured for \pks\ from \chandra\ and \xmm\ observations. \label{fig:gamma}}
\end{figure}

\begin{figure}
\epsscale{1}
\plotone{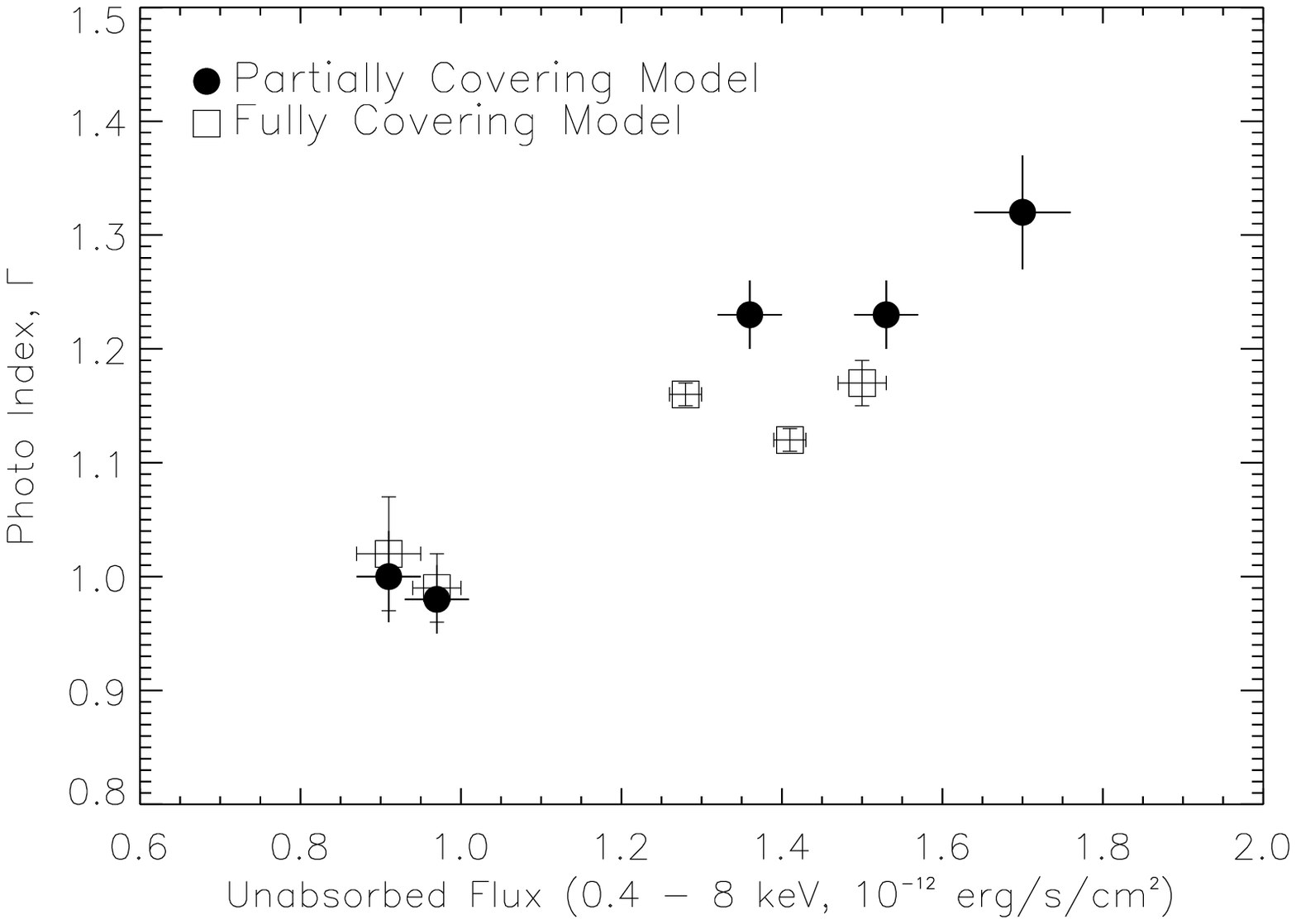}
\caption{The photon indices versus flux for \pks\ from \chandra\ and \xmm\ observations. \label{fig:gammaf}}
\end{figure}

We detected the variable absorption, variable power-law photon index, and
variable flux for the blazar \pks.  
The observed variability between \chandra\ and \xmm\ spectra could be,
in part, due to imperfect cross-calibration between the two.
However for high S/N spectra, the 
cross-calibration between \chandra\ and \xmm\ can
yield a spectral difference of $\Delta \Gamma \sim$ 0.03\footnote{http://xmm.esac.esa.int/external/xmm\_sw\_cal/calib/cross\_cal/index.php}, 
whereas our measured differences are between $\Delta \Gamma \sim$ 0.15--0.3.
In addition, the higher S/N \xmm\ spectra show a statistically significant variability 
within the three \xmm\ observations ($\Delta \chi^2$ improvement of 14.5, 14.7,
and 14.1 for Models 3, 4, and 5 with spectral variability).
We also found the spectral variability between the two \chandra\ observations, 
where the second \chandra\ epoch has consistently lower absorption column densities than the first observation.  This is better demonstrated in \S~3.1, where we analyzed the individual spectra of each image.
We plot the variations of absorption
in Figure~\ref{fig:nh}, photon index variations in
Figure~\ref{fig:gamma}, and the photon index against unabsorbed flux in
Figure~\ref{fig:gammaf}.  It is well known that blazars are variable,
so the observed spectral variability is not surprising in itself.
We also found that the photon index and flux are correlated
(Figure~\ref{fig:gammaf}).  This relation has been observed in several
Gamma-ray loud AGNs (Foschini et al. 2006).  \pks\ provides an
additional example of a source following this correlation.

The variable absorption is more intriguing.  
The variable absorption toward \pks\ was 
detected previously with \asca\ observations \citep{oshima01}, although 
interpreted differently.  \citet{oshima01} obtained a variation
amplitude of $\sim 0.5\times10^{22} \cmsq$ when modeling the absorption
at the lens, which is consistent with our measurement of
$\sim0.6\times10^{22} \cmsq$ (Model 3).  We note that the variation
detected by \citet{oshima01} is on the timescale of 10 days while that
found by us is on the timescale of years.  It is possible that the
variable absorption is at the source redshift.  Depending on the nature
of the absorber, we obtained variations of $\sim 2\times10^{22} \cmsq$
and $\sim 30\times10^{22} \cmsq$ for a fully covering absorber (Model 4) and a
partially covering absorber (Model 5, 6), respectively.  The other
possibilities are microlensing models.

\citet{oshima01} proposed a microlensing interpretation to explain the
 apparent variable absorption.  The basic idea is that the absorption
 does not vary but is different for the two images, and the microlensing
 produces flux changes between the two images which appears as variable
 absorption.  As we discussed in the spectral fits to individual
 \chandra\ images (\S~\ref{sec:indi}), the absorption varies in
 individual images between the two epochs, which cannot be attributed to
 this simple microlensing model.  For this reason, we rule out the
 simple microlensing model as the cause of variable absorption in \pks.
 More complicated microlensing models are required which must involve
 significant emission size differences between the soft and hard X-ray
 regions to interpret the data.  However, even these models will have
 difficulty explaining the similar differential absorptions between
 images A and B for the two \chandra\ epochs.  Microlensing will only
 produce uncorrelated changes between the images.

Variable absorption thus appears to be a robust result, independent of
model.  The variability can either come from the lens or from the source
redshift.  If the variable absorption is at the lens, a variability
amplitude of $\sim0.6\times10^{22} \cmsq$ can sufficiently parameterize
the \chandra\ and \xmm\ spectra without invoking any absorption at the
source.  Recently, variable Mg~II absorption has been detected in an
intervening absorber in front of a Gamma-ray burst \citep{hao06}.  The
observed Mg~II column density changes as the burst expands geometrically
\citep{frank06}.  Similarly, if the size of the blazar changes with
luminosity, it may produce variable absorption at the lens.  Another
mechanism that can produce variable absorption at the lens galaxy is
proposed by \citet{dong06}, where microlensed images, slightly shifted
from the original image position, intersect slightly different positions
at the lens, producing variable absorption.  However, in \pks, the
differential absorption of the two images in the two \chandra\ epochs is
similar, arguing against the variability at the lens. It is far more likely
that the variable absorption is intrinsic to the blazar, simply because
there are other AGNs with variable intrinsic absorption, including
blazars (see below).

While the absorption toward the gravitational lens \pks\ is very complicated,
with the sequence of \chandra\ and \xmm\ observations, a converging
picture is emerging.  Besides the Galactic absorption of $\nh =
22.27\times10^{20}~\cmsq$ \citep{d90}, there is also an intrinsic
absorption and absorption from the lens galaxy.  The absorption column
densities in the lens galaxy are different for the two lensed images by
$\Delta\nh_{B,A} = 1.8^{+0.5}_{-0.6}\times 10^{22}$~\cmsq\
\citep{dai06}, and this is consistent with the differential extinction
measurement between the two images and a Galactic dust-to-gas ratio.
The amplitude of the variability of the intrinsic absorption is model
dependent with amplitudes of $\sim 2\times10^{22}$--$30 \times10^{22} \cmsq$.

Recently, the intrinsic absorption for blazars has been reported in several
cases, such as GB~B1428+4217 \citep{worsley04a}, PMN~J0525$-$3343
\citep{worsley04b}, and RBS~315 (Piconcelli \& Guainazzi 2005; see also Tavecchio et al. 2007).  In this paper, we present
variable intrinsic absorption toward gravitationally-lensed
blazar \pks.  The large variation in the absorption column density can be
interpreted as outflows similar to those detected in BAL~QSOs.  However,
as \pks\ is a blazar, this outflow must be in the polar
direction. Recent numerical simulations of accretion flows in AGNs
\citep{proga06} discussed the cases of polar outflows; our observations of
\pks\ might be providing an example in support of such models.
In addition, it is also possible that the intrinsic absorption variation
is caused by the interaction between the blazar jet and its environment,
or the variation from the geometric configuration of the jet.
More intense monitoring of this system or further 
variability studies using a large sample of blazars are needed to better
constrain the nature of their X-ray spectral variation.


\clearpage

\begin{deluxetable}{cccc}
\tabletypesize{\scriptsize}
\tablecolumns{4}
\tablewidth{0pt}
\tablecaption{List of X-Ray Observations of \pks\label{tab:obslog}}
\tablehead{
\colhead{Date} &
\colhead{Telescope} &
\colhead{Grating?} &
\colhead{Exposure Time (sec)} 
}
\startdata
2000-06-26 & \chandra & HETGS & 47471 \\
2001-06-25 & \chandra & HETGS & 51219 \\
2004-03-10 & \xmm     & none  & 8328  \\
2004-03-24 & \xmm     & none  & 27004 \\
2004-05-05 & \xmm     & none  & 21356 \\
\enddata
\end{deluxetable}

\begin{deluxetable}{ccccc}
\tabletypesize{\scriptsize}
\tablecolumns{5}
\tablewidth{0pt}
\tablecaption{Spectral Analysis for Individual \chandra\ Spectra of \pks.\label{tab:indi}}
\tablehead{
\colhead{Date} &
\colhead{Image} &
\colhead{\nh (z=0.886), $10^{22}$~\cmsq} &
\colhead{$\Gamma$} &
\colhead{$\chi^{2}(dof)$}
}

\startdata
\cutinhead{Model 1: An absorbed power-law} 
2000-06-26 & A & $1.6\pm0.4$ & $1.09\pm0.09$ & 100.5(108) \\
2000-06-26 & B & $3.5\pm1.1$ & $1.13\pm0.15$ & 129.6(106) \\
2001-06-25 & A & $0.8\pm0.5$ & $0.96\pm0.09$ & 140.7(120) \\
2001-06-25 & B & $3.3\pm1.0$ & $1.11\pm0.12$ & 113.8(116) \\
\cutinhead{Model 2: An absorbed power-law with $\Gamma_A = \Gamma_B$.} 
2000-06-26 & A & $1.6\pm0.4$ & $1.10\pm0.07$ & 229.6(215) \\
2000-06-26 & B & $3.4\pm0.7$ & \nodata       & \nodata    \\
2001-06-25 & A & $1.0\pm0.4$ & $1.02\pm0.06$ & 255.2(237) \\
2001-06-25 & B & $2.8\pm0.6$ & \nodata       & \nodata    \\
\enddata

\tablecomments{Galactic \nh\ is fixed at $\nh = 22.27\times10^{20}~\cmsq$ (Dickey \& Lockman 1990) for all models.}
\end{deluxetable}

\clearpage

\begin{deluxetable}{lcccccc}
\tabletypesize{\scriptsize}
\tablecolumns{7}
\tablewidth{0pt}
\tablecaption{Spectral Analysis of Combined Images of \pks.\label{tab:spec}}
\tablehead{
\colhead{} &
\multicolumn{2}{c}{\chandra} &
\multicolumn{4}{c}{\xmm} \\
\cline{2-3} 
\cline{4-7}
\colhead{Parameters} &
\colhead{I} &
\colhead{II} &
\colhead{I}  &
\colhead{II}  &
\colhead{III} &
\colhead{XMM-Co-Added}
}

\startdata
Date & 2000-06-26 & 2001-06-25 & 2004-03-10 & 2004-03-24 & 2004-05-05 & \nodata \\
\cutinhead{Model 3: A power-law with absorption at the lens and the Galaxy} 
\nh ($z=0.89, 10^{22}~\cmsq$)    & $2.27\pm0.26$ & $1.70\pm0.22$ & $2.42\pm0.12$ & $2.37\pm0.07$ & $2.29\pm0.07$ & $2.34\pm0.04$ \\
$\Gamma$                         & $1.07\pm0.05$ & $0.99\pm0.04$ & $1.18\pm0.02$ & $1.14\pm0.01$ & $1.17\pm0.01$ & $1.16\pm0.01$ \\
Unabsorbed Flux \tablenotemark{a} & $0.92\pm0.02$ & $0.97\pm0.03$ & $1.50\pm0.03$ & $1.42\pm0.02$ & $1.28\pm0.02$ & $1.38\pm0.01$ \\
$\chi^{2}(dof)$                  & 365.6(305)    & 365.6(337)    & 151.8(152)    & 447.3(396)    & 314.3(301)    & 843.7(743)    \\
\cutinhead{Model 4: Same as Model 3, with additional absorption at the source\tablenotemark{b}}
\nh ($z=2.51, 10^{22}~\cmsq$)    & $1.24^{+1.10}_{-1.19}$ & $<0.38$       & $3.47\pm0.60$ & $2.80\pm0.33$ & $2.68\pm0.36$ & $2.80\pm0.22$ \\
$\Gamma$                         & $1.02\pm0.05$          & $0.99\pm0.03$ & $1.17\pm0.02$ & $1.12\pm0.01$ & $1.16\pm0.01$ & $1.14\pm0.01$ \\
Unabsorbed Flux               & $0.91\pm0.04$             & $0.97\pm0.03$ & $1.50\pm0.03$ & $1.41\pm0.02$ & $1.28\pm0.02$ & $1.38\pm0.01$ \\
$\chi^{2}(dof)$                  & 369.9(305)             & 365.6(337)    & 157.6(152)    & 484.5(396)    & 327.3(301) & 899.3(743) \\
\cutinhead{Model 5: Same as Model 4, but with a partially covering absorber at the source\tablenotemark{b}}
\nh ($z=2.51, 10^{22}~\cmsq$)    & $<4.6$        & $<0.7$       & $29.0^{+8.0}_{-4.8}$  & $24.2^{+4.3}_{-2.8}$  & $19.5^{+6.1}_{-3.8}$  & $25.0\pm2.5$  \\
Covering Factor                  & 0.44 (fixed)  & 0.44 (fixed)  & $0.50\pm0.05$ & $0.43\pm0.03$ & $0.40\pm0.04$ & $0.44\pm0.02$ \\
$\Gamma$                         & $1.00\pm0.04$ & $0.98\pm0.03$ & $1.32\pm0.05$ & $1.23\pm0.03$ & $1.23\pm0.03$ & $1.26\pm0.02$ \\
Unabsorbed Flux                  & $0.91\pm0.04$ & $0.97\pm0.04$ & $1.70\pm0.06$ & $1.53\pm0.04$ & $1.36\pm0.04$ & $1.51\pm0.03$ \\
$\chi^{2}(dof)$                  & 370.6(305)    & 365.9(337)    & 137.4(151)    & 447.4(395)    & 315.6(300)    & 827.0(742)    \\
\cutinhead{Model 6: Same as Model 5, but with two different power-law components\tablenotemark{c}}
$\nh_A$ ($z=0.89, 10^{22}~\cmsq$) & \multicolumn{6}{c}{$1.57\pm0.07$} \\
\nh ($z=2.51, 10^{22}~\cmsq$)     & $<1.7$ & $<0.7$ & \nodata & \nodata & \nodata & $39\pm7$ \\
Covering Factor                   & \multicolumn{6}{c}{$0.23\pm0.04$} \\
$\Gamma$                          & $1.07\pm0.03$ & $1.07\pm0.02$ & \nodata & \nodata & \nodata & $1.27\pm0.03$ \\
$\chi^{2}(dof)$                   & \multicolumn{6}{c}{1488.8(1367)} \\
\enddata

\tablecomments{Galactic \nh\ is fixed at $\nh = 22.27\times10^{20}~\cmsq$ (Dickey \& Lockman 1990) for all models.}
\tablenotetext{a}{The unabsorbed fluxes in all models are calculated between 0.4--8 keV and in units of $10^{-12}$\flux.}
\tablenotetext{b}{The absorption at lens is fixed at $\nh = 1.7\times10^{22}~\cmsq$ for this model.}
\tablenotetext{d}{The absorption of the lens galaxy for image B is constrained as $\nh_B = \nh_A + 1.8\times10^{22}~\cmsq$.}
\end{deluxetable}

\clearpage

\clearpage

\clearpage

\clearpage

\end{document}